\documentclass[prd,aps,
twocolumn,
nofootinbib,
preprintnumbers,
superscriptaddress]{revtex4}
\usepackage[T1]{fontenc} % if needed
\usepackage{graphicx}
\usepackage{epsfig}
\usepackage{xcolor}
\usepackage{rotating}
\usepackage{amssymb}
\usepackage{subfigure}
\usepackage{dsfont}
\usepackage{psfrag}
\usepackage{amsmath,euscript,array,mathrsfs}
\usepackage{bbold}
\usepackage{epsf}
\usepackage[utf8]{inputenc}
\usepackage{placeins}
\usepackage{dcolumn}

\newcommand{\beq}{\begin{equation}}
\newcommand{\eeq}{\end{equation}}
\newcommand{\beqs}{\begin{eqnarray}}
\newcommand{\eeqs}{\end{eqnarray}}

\newcommand{\gsim}{\mathrel{\raisebox{-.6ex}{$\stackrel{\textstyle>}{\sim}$}}}
\newcommand{\Tr}{{\rm Tr}}

\def\hbar{\hspace{0pt}\raisebox{1pt}{$-$} \hspace{-7pt} h}

\def\di{\mbox{d}}

\newcommand{\be}{\begin{equation}}
\newcommand{\ee}{\end{equation}}

\newcommand{\bea}{\begin{eqnarray}}
\newcommand{\eea}{\end{eqnarray}}

\def\lbldef#1#2{\expandafter\gdef\csname #1\endcsname {#2}}

\def\href#1#2{#2}

\newcommand{\ber}{\begin{eqnarray}}
\newcommand{\eer}{\end{eqnarray}}

\newcommand{\beqar}{\begin{eqnarray}}

\newcommand{\eeqar}{\end{eqnarray}}

\newcommand{\dsl}
  {\kern.06em\hbox{\raise.15ex\hbox{$/$}\kern-.56em\hbox{$\partial$}}}

\newcommand{\eeqarr}{\end{eqnarray}}
\newcommand{\ZZ}{{\rm \kern 0.275em Z \kern -0.92em Z}\;}

\def\CC{{\mathchoice
{\rm C\mkern-8mu\vrule height1.45ex depth-.05ex
width.05em\mkern9mu\kern-.05em}
{\rm C\mkern-8mu\vrule height1.45ex depth-.05ex
width.05em\mkern9mu\kern-.05em}
{\rm C\mkern-8mu\vrule height1ex depth-.07ex
width.035em\mkern9mu\kern-.035em}
{\rm C\mkern-8mu\vrule height.65ex depth-.1ex
width.025em\mkern8mu\kern-.025em}}}

\def\RR{{\rm I\kern-1.6pt {\rm R}}}

\def\ZZ{{\rm Z}\kern-3.8pt {\rm Z} \kern2pt}
\def\IB{\relax{\rm I\kern-.18em B}}
\def\ID{\relax{\rm I\kern-.18em D}}
\def\II{\relax{\rm I\kern-.18em I}}
\def\IP{\relax{\rm I\kern-.18em P}}

\newcommand{\bear}{\begin{eqnarray}}
\newcommand{\eear}{\end{eqnarray}}

\def\to{\rightarrow}

\def\to{\rightarrow}

\def\i{\iota}

\def\6{\partial}

\def\bea{\begin{eqnarray}}
\def\eea{\end{eqnarray}}

\def\beqx{\begin{displaymath}}
\def\eeqx{\end{displaymath}}

\newcommand{\bmat}{\left(\begin{array}}
\newcommand{\emat}{\end{array}\right)}

\def\i{\iota}

\def\bo{{\raise-.3ex\hbox{\large$\Box$}}}               % D'Alembertian
\def\face{{\raise.2ex\hbox{$\displaystyle \bigodot$}\mskip-2.2mu \llap {$\ddot \smile$}}}% happy face
\def\>{\rangle}                                      %right angle
\def\<{\langle}                                      %left angle

\def\leftrightarrowfill{$\mathsurround=0pt \mathord\leftarrow \mkern-6mu
        \cleaders\hbox{$\mkern-2mu \mathord- \mkern-2mu$}\hfill
        \mkern-6mu \mathord\rightarrow$}        % <--> double differential
\def\dvec#1{\vbox{\ialign{##\crcr
        \leftrightarrowfill\crcr\noalign{\kern-1pt\nointerlineskip}
        $\hfil\displaystyle{#1}\hfil$\crcr}}}           % <--> accent
\def\Tr{{\rm Tr \,}}                                    % Trace
\def\-{\hphantom{-}}

\begin{document}
\title{Color dependence of the topological susceptibility in Yang-Mills theories}

\author{Ed Bennett}
\email{e.j.bennett@swansea.ac.uk}
\affiliation{Swansea Academy of Advanced Computing, Swansea University,
Fabian Way, SA1 8EN, Swansea, Wales, UK}
\author{Deog Ki Hong}
\email{dkhong@pusan.ac.kr}
\affiliation{Department of Physics, Pusan National University, Busan 46241, Korea}
\author{Jong-Wan Lee}
\email{jwlee823@pusan.ac.kr}
\affiliation{Department of Physics, Pusan National University, Busan 46241, Korea}
\affiliation{Institute for Extreme Physics, Pusan National University, Busan 46241, Korea}
\author{C.-J.~David~Lin}
\email{dlin@nycu.edu.tw}
\affiliation{Institute of Physics, National Yang Ming Chiao Tung University, 1001 Ta-Hsueh Road, Hsinchu 30010, Taiwan}
\affiliation{Center for High Energy Physics, Chung-Yuan Christian University,
Chung-Li 32023, Taiwan}
\affiliation{Centre for Theoretical and Computational Physics, National Yang Ming Chiao Tung University, 1001 Ta-Hsueh Road, Hsinchu 30010, Taiwan}
\affiliation{Physics Division, National Centre for Theoretical Sciences, Taipei 10617, Taiwan}
\author{Biagio Lucini}
\email{b.lucini@swansea.ac.uk}
\affiliation{Department of Mathematics, Faculty  of Science and Engineering,
Swansea University, Fabian Way, SA1 8EN Swansea, Wales, UK}
\affiliation{Swansea Academy of Advanced Computing, Swansea University,
Fabian Way, SA1 8EN, Swansea, Wales, UK}
\author{Maurizio Piai}
\email{m.piai@swansea.ac.uk}
\affiliation{Department of Physics, Faculty  of Science and Engineering,
Swansea University,
Singleton Park, SA2 8PP, Swansea, Wales, UK}
\author{Davide Vadacchino}
\email{davide.vadacchino@plymouth.ac.uk}
\affiliation{School of Mathematics and Hamilton Mathematics Institute, Trinity
College, Dublin 2, Ireland}
\affiliation{Centre for Mathematical Sciences, University of Plymouth, Plymouth, PL4 8AA, United Kingdom}

%\date{\today}

\begin{abstract}

For Yang-Mills theories in four dimensions,
we propose  to rescale  the 
ratio between topological susceptibility
and string tension squared in a universal way,
 dependent only on
group factors. We apply this suggestion to 
$SU(N_c)$ and $Sp(N_c)$ groups,
and compare  lattice measurements  
performed by several independent collaborations.
We show that the two sequences of (rescaled) numerical results in
 these two families
of groups are compatible with each other.
We hence  perform a combined fit, and
 extrapolate
to the common  large-$N_c$ limit.

\end{abstract}

\maketitle
%\tableofcontents
\preprint{PNUTP-22/A02}

%%%%%%%%%%%%%%%%%%%%%
%%%%%%%%%%%%%%%%%%%%%
\section{Introduction}
\label{Sec:introduction}

Lattice studies provide numerical evidence that,  at zero temperature,
four-dimensional Yang-Mills theories 
with compact non-Abelian gauge group $G$ confine.
This statement can be made precise, for instance 
by formulating it in terms of the expectation values of 
either the Polyakov loop or the Wilson loop,
and then extracting the string tension $\sigma$ from suitable correlation functions.
It is of general interest to identify other observables
that 
characterise the long-distance behaviour of Yang-Mills theories, 
for all choices of group $G$.
By doing so, one can relate lattice results to
alternative approaches  based on the large-$N_c$ expansion.
A resurgence of interest in the latter, motivated by
 gauge-gravity dualities~\cite{Maldacena:1997re,
Gubser:1998bc,Witten:1998qj,Aharony:1999ti}, led to
much effort being focused on the
 glueballs, as the results of
lattice calculations of their spectra~\cite{Lucini:2001ej,
Lucini:2004my,Lucini:2010nv,Lucini:2012gg,Athenodorou:2015nba,
Lau:2017aom,Bennett:2017kga,Bennett:2020hqd,Bennett:2020qtj,Hernandez:2020tbc,
Yamanaka:2021xqh,Athenodorou:2021qvs,Bonanno:2022yjr}
can be compared to those of gravity 
calculations~\cite{Brower:2000rp,Apreda:2003sy,
Mueck:2004qg,Wen:2004qh,Kuperstein:2004yf,
Elander:2013jqa,Athenodorou:2016ndx,
Elander:2018aub,Elander:2020csd,
Elander:2021kxk}---or
 other semi-analytical calculations~\cite{Hong:2017suj,Bochicchio:2016toi,Bochicchio:2013sra}.

The topological susceptibility, $\chi$, is a non-perturbative quantity that
 plays a central role in our understanding of
strong nuclear forces---see for instance the review in Ref.~\cite{Vicari:2008jw}. 
It enters the Witten-Veneziano formula~\cite{Witten:1979vv,Veneziano:1979ec} 
for the mass of the $\eta^{\prime}$ particle,
and the solution of the $U(1)_A$ problem. Being related to the $\theta$-dependence of the free energy,
$\chi$ also enters the electric dipole moment of hadrons,
the strong-CP problem, and its putative solutions (the
axion).
Being topological in nature, $\chi$ is intrinsically difficult to
compute on the lattice; yet, modern lattice techniques are mature enough that increasingly precise and 
reliable measurements have been published 
 in the past two decades for  $SU(N_c)$ Yang-Mills 
 theories~\cite{Lucini:2001ej,DelDebbio:2002xa,Bonati:2016tvi,Bonanno:2020hht,Athenodorou:2021qvs}---see also
 Refs.~\cite{Lucini:2004yh,DelDebbio:2004ns,Panagopoulos:2011rb, Bonati:2015sqt,
Ce:2016awn,Borsanyi:2021gqg,Luscher:1981zq,Campostrini:1989dh,Luscher:2010ik,
 Luscher:2011kk,Alexandrou:2017hqw,Cossu:2021bgn,Teper:2022mmj}.
Our collaboration has just completed the calculation of $\chi$ in the $Sp(N_c)$ Yang-Mills
theories~\cite{Bennett:2022ftz}.
In this paper we propose a way to compare $\chi$ in different
sequences of gauge groups, and  perform a combined large-$N_c$ extrapolation.

%%%%%%%%%%%%%%%%%%%%
%%%%%%%%%%%%%%%%%%%%
\section{Yang-Mills theories}
\label{Sec:theory}

The Yang-Mills theory with gauge group $G$, 
in four-dimensional Minkowski space, has the classical action:
\beqs
{\cal S}_{YM}&=&-\frac{1}{2g^2}\int \di^4 x \,\Tr F_{\mu\nu}F^{\mu\nu}\,,
\label{Eq:YM}
\eeqs
with $g$ the coupling,  $F_{\mu\nu}\equiv \partial_{\mu}
 A_{\nu} - \partial_{\nu} A_{\mu}+i [A_{\mu}\,,\,A_{\nu}]$
 the field-strength tensor,
 and $A_{\mu}\equiv \sum_AA_{\mu}^AT^A$  the 
 gauge field. The matrices $T^A$,  with $A=1,\,\cdots,\,d_G$, 
are the generators  in the fundamental representation, 
 normalised by the relation  $\Tr T^AT^B =\frac{1}{2}\delta^{AB}$.

Yang-Mills theories are asymptotically free at short distance,
hence can be interpreted as conformal theories 
admitting  a marginally relevant deformation: the gauge coupling.
Long distance physics is not accessible to perturbative calculations;
its numerical treatment is implemented by discretising the Euclidean spacetime on a lattice.
The discretised action and  
range of its parameters are chosen
so that Monte Carlo numerical studies are performed
within the basin of attraction of a fixed point
belonging to the universality class of the aforementioned conformal theory.
By doing so, it is possible to suppress 
non-universal features of the lattice formulation 
and study the universal properties of the gauge dynamics characterising
 the continuum, four-dimensional physical system of interest.
Observable quantities are measured as 
ensemble averages of appropriately chosen operators, and
 extrapolated towards the continuum limit,
  where the lattice spacing $a$ vanishes,
by changing the lattice parameters 
so as to approach the fixed point in a controlled way.

We do not report the details of the lattice theories of interest here, 
except for highlighting the fact that
in comparing  measurements with different ensembles,
and extrapolating towards the continuum limit, 
one measures the dimensional observables of interest in units of a physical scale, 
hence introducing a  scale setting procedure.
We compare measurements in different theories, performed by different collaborations, with different
lattice algorithms, but all of them adopting the same scale-setting procedure,
 based upon the string tension $\sigma$.

%%%%%%%%%%%%%%%%%%%%
%%%%%%%%%%%%%%%%%%%%
\subsection{String Tension}
\label{Sec:string}

On the lattice, to extract  the
string tension $\sigma$ one measures the
correlation
functions between non-contractible path-ordered loops,
separated by Euclidean distance $L$.
The resulting fluxtubes are 
described by effective string theory  when $L/a\gg 1$, 
and the mass $a m(L)$ (in lattice units) of the
lightest (torelon) state is
\begin{eqnarray}
a m (L) &=& (\sigma a^2) \frac{L}{a}\left(1+\sum_{k=1}^{+\infty}
\frac{d_k}{(\sigma L^2)^k}\right)\,.
\end{eqnarray}
The effective string theory~\cite{Aharony:2009gg}
is characterised by the values of $d_k$,
dimensionless coefficients
that capture the dynamics at large distances;
$d_1=-\pi/3$ is the universal L\"uscher
term~\cite{Luscher:1980ac}.
One  estimates $\sigma a^2$
by repeating lattice measurements for different $L/a$, 
and curve-fitting the results.
For further details on the measurements of $\sigma a^2$, we refer the reader to Ref.~\cite{Bennett:2020hqd}, for example.

Lattice  measurements are affected by
both  statistical and systematic uncertainties that 
are difficult to reduce below the few percent level.
Furthermore, one intrinsic limiting factor in the adoption of $\sigma$
as a universal scale setting procedure in non-Abelian gauge theories  is that
$\sigma$ is not well defined for asymptotically large $L$,
if  string-breaking effects are present, 
as is the case with  dynamical matter fields.
Yet  many lattice collaborations
report their results in terms of $\sigma$, because of the simplicity 
of its extraction and its intuitive meaning.
We  adopt this strategy for the purposes of this paper,
and in this work we do not attempt to compare with results that 
use a different scale setting method, such as the gradient flow, as done, e.g., in Ref.~\cite{Ce:2016awn}. 
\\

%%%%%%%%%%%%%%%%%%%%
%%%%%%%%%%%%%%%%%%%%
\subsection{Topological Susceptibility}
\label{Sec:susceptibility}

The topological charge $Q$ of a gauge configuration is
\begin{equation}
    Q \equiv \int \mathrm{d}^4 x~q(x)\,,
\end{equation}
where
\begin{equation}\label{eq:cont_top_charge_density}
    q(x) \equiv \frac{1}{32\pi^2}
    \varepsilon^{\mu\nu\rho\sigma}\,
    \mathrm{Tr}~F_{\mu\nu}(x) F_{\rho\sigma}(x)\,,
\end{equation}
 with $\varepsilon^{\mu\nu\rho\sigma}$  is the  Levi-Civita symbol.
The topological susceptibility is defined as
\begin{equation}
    \chi \equiv \int \mathrm{d}^4 x ~\langle q(x) q(0) \rangle.
\end{equation}
The inclusion of a $\theta$ term yields the action $\tilde{\cal S}$, which extends 
 Eq.~(\ref{Eq:YM}):
 \begin{widetext}
 \beqs
 \label{Eq:Stilde}
\tilde{{\mathcal{S}}} &=&
 -\frac{1}{2g^2}\int \di^4 x \,\Tr F_{\mu\nu}F^{\mu\nu}
 -\frac{\theta}{32\pi^2}\int \di^4 x \,\varepsilon^{\mu\nu\rho\sigma}\Tr F_{\mu\nu}F_{\rho\sigma}\,.
 \eeqs
 \end{widetext}
The vacuum (free) energy (density) $F(\theta)$  is defined by the path integral
\beqs
e^{-V_4 F(\theta)}&\equiv& \int {\cal D}A_{\mu} e^{-\tilde{\cal S}_E}\,,
\eeqs
where $V_4$ is the four-dimensional volume, and $\tilde{\cal S}_E$ the Euclidean version of Eq.~(\ref{Eq:Stilde}).
The topological susceptibility is then computed as
\beqs
\chi&=&\left.\frac{\partial^2 F(\theta)}{\partial \theta^2}\right|_{\theta=0}\,.
\eeqs

In the continuum theory, the charge  $Q\in \mathbb{Z}$ is quantised.
Lattice artefacts spoil the discreteness of the topological charge and prevent $Q$ from taking integer values 
on configurations generated in numerical simulations. The assignment of integer topological charge 
on the lattice is affected by an ambiguity, though this is expected to be irrelevant in the continuum limit. 

Other factors that affect the accuracy of the results stem from the practical limitations of Monte Carlo updating algorithms and of the finite range of lattice spacings that can be simulated. Among them, we mention the existence of
(auto)correlation between configurations, (partial)
topological freezing, and numerical noise due to 
short-distance fluctuations, as well as the appearance of other uncertainties in the
continuum limit extrapolation.
We refer to the original literature for details~\cite{Lucini:2001ej,DelDebbio:2002xa,
Bonati:2016tvi,Bonanno:2020hht,
Athenodorou:2021qvs,Bennett:2022ftz,Panagopoulos:2011rb, Bonati:2015sqt},
and for a survey of the advanced strategies that the lattice collaborations implement in order to minimise the statistical error and the systematic effects in the measurement of $\chi$. Under the reasonable assumption that the identified errors have been evaluated correctly, a direct comparison of the results from
the measurements of the different groups is a way to assess the size of any potentially remaining systematic effects.

%%%%%%%%%%%%%%%%%%%%
%%%%%%%%%%%%%%%%%%%%
\section{Towards large $N_c$}
\label{Sec:Nc}

Since the $\theta$ term is topological, it does not affect the local dynamics of the gauge
 fields, such as the running coupling. It is therefore widely believed that at low energy
Yang-Mills theories 
confine even 
in the presence of a non-vanishing $\theta$,  at least as 
long as $\theta$ is small. 
The $\theta$-dependent vacuum  is gapped, and all  the excitations (glueballs)
are color-singlets. 
In order for CP to be a well defined symmetry, we also expect
the vacuum energy to be an even function of $\theta$,  minimised
at $\theta=0$, by consequence of
 the Schwarz inequality applied to the Euclidean partition function~\cite{Vafa:1984xg,Vafa:1983tf}:
\begin{equation}
F(0)\le F(\theta)=F(-\theta)\,.
\end{equation}

By defining the 't Hooft coupling $\lambda\equiv g^2 N_c$, because the trace of any $N_c\times N_c$
 matrix is proportional to $N_c$, while the couplings are proportional to $\lambda/N_c$,
Yang-Mills theories can be analysed in a $1/N_c$ expansion in which one holds $\lambda$ fixed. 
For consistency at the quantum level,
the $\theta$ term must be scaled holding $\theta/N_c$  fixed as well,
and physical observables are multi-valued 
functions of $\theta$ with periodicity $2\pi $~\cite{Witten:1998uka}. 
For example, the vacuum energy is expected to take the form
\begin{equation}
F(\theta)=f_G\,\underset{k}{\rm min}\,h\left(\frac{\theta+2\pi k}{N_c}\right)\,,
\end{equation}
with $k=0,\cdots, N_c-1$, and the pre-factor  $f_G={\cal O}\left(N_c^2\right)$ for large $N_c$.
$h$ is smoothly 
dependent on $\theta/N_c$ for small $\theta$, and is determined by $G$
in a way that admits a finite limit as $N_c\rightarrow \infty$.
For $\theta=0$,
the minimum is expected for $k=0$~\cite{Witten:1998uka}, and
the large-$N_c$ limit of the topological susceptibility is finite:
\begin{equation}
\lim_{N_c\rightarrow \infty}\chi=\chi_{\infty}\,,
\end{equation}
with $\chi_{\infty}=h''(0)$.  
As each gauge field contributes equally, one expects that
\begin{equation}
f_G\propto d_G\,,
\end{equation}
where $d_G$ is the dimension of the group;  $d_G=N_c^2-1$ for $SU(N_c)$ 
and $d_G=(N_c+1)N_c/2$ for $Sp(N_c)$. The proportionality factor
must be finite in the large-$N_c$ limit.

The string tension is the energy density per unit length of a fluxtube,
the limiting case of a fermion-antifermion pair in the fundamental representation,
separated by an asymptotically large distance.
We hence expect $\sigma$ to be 
 proportional to the strength of the coupling between the fermions, which
 can be measured by the
 quadratic Casimir of the fundamental representation~\cite{Hong:2017suj}:
\begin{equation}
\sigma\propto C_2(F)\,=\,
\begin{cases}
\frac{N_c^2-1}{2N_c} & \text{ for } {SU}(N_c) \\
\frac{N_c+1}{4} & \text{ for } {Sp}(N_c)
\end{cases}\,.
\end{equation}
The proportionality factor is itself a function of $N_c$, and encodes non-perturbative dynamics in such a way that
the string tension has a finite large-$N_c$ limit, $\sigma_{\infty}$, as expected
because the coupling of fundamental 
fermions scales as $1/\sqrt{N_c}$,
  while there are $N_c$ components to them.

The topological susceptibility inherits its 
group-dependence from the vacuum energy. 
Hence, we expect the following ratio to capture universal features:
\begin{equation}
\eta_{\chi}\equiv\frac{\chi C_2(F)^2}{\sigma^2 d_G} \,
=\frac{\chi}{\sigma^2}\cdot 
\begin{cases}
\frac{N_c^2-1}{4N_c^2} & \text{ for } {SU}(N_c) \\
\frac{N_c+1}{8N_c} & \text{ for } {Sp}(N_c)
\end{cases}\,.
\end{equation}
Furthermore, we expect the ratio $\eta_{\chi}$ 
  to be finite and universal in the limit $N_c\to\infty$:
\begin{equation}
\lim_{N_c\to\infty}\frac{\chi C_2(F)^2}{\sigma^2 d_G}=b\,\frac{\chi_{\infty}}{\sigma_{\infty}^2}=\eta_{\chi}(\infty) <\infty\,,
\end{equation}
where $b=1/4$ for $SU(N_c)$, while $b=1/8$ for $Sp(N_c)$.

%%%%%%%%%%%%%%%%%%%%
%%%%%%%%%%%%%%%%%%%%
\section{Numerical Results}
\label{Sec:numbers}

\begin{table}

\caption{Summary table of measurements used in this
study.\label{Fig:table}\\}
\centering
\begin{tabular}{|c|c|c|c|}
\hline\hline
Group & Reference & $\chi/\sigma^2$ & $C_2(F)^2/d_G$ \\
\hline
$Sp(2)$ & Bennett et al.~\cite{Bennett:2022ftz} & $ 0.0519(27) $ & $ 0.1875 $ \\
$Sp(4)$ & Bennett et al.~\cite{Bennett:2022ftz} & $ 0.0424(27) $ & $ 0.1562 $ \\
$Sp(6)$ & Bennett et al.~\cite{Bennett:2022ftz} & $ 0.0396(49) $ & $ 0.1458 $ \\
$Sp(8)$ & Bennett et al.~\cite{Bennett:2022ftz} & $ 0.0424(40) $ & $ 0.1406 $ \\
\hline
$SU(2)$ & Lucini et al.~\cite{Lucini:2001ej} & $ 0.0507(24) $ & $ 0.1875 $ \\
$SU(3)$ & Lucini et al.~\cite{Lucini:2001ej} & $ 0.0355(32) $ & $ 0.2222 $ \\
$SU(4)$ & Lucini et al.~\cite{Lucini:2001ej} & $ 0.0224(39) $ & $ 0.2344 $ \\
$SU(5)$ & Lucini et al.~\cite{Lucini:2001ej} & $ 0.0224(49) $ & $ 0.2400 $ \\
\hline
$SU(3)$ & Del Debbio et al.~\cite{DelDebbio:2002xa} & $ 0.0282(12) $ & $ 0.2222 $ \\
$SU(4)$ & Del Debbio et al.~\cite{DelDebbio:2002xa} & $ 0.0257(10) $ & $ 0.2344 $ \\
$SU(6)$ & Del Debbio et al.~\cite{DelDebbio:2002xa} & $ 0.0236(10) $ & $ 0.2431 $ \\
\hline
$SU(4)$ & Bonati et al.~\cite{Bonati:2016tvi} & $ 0.02480(80) $ & $ 0.2344 $ \\
$SU(6)$ & Bonati et al.~\cite{Bonati:2016tvi} & $ 0.02300(80) $ & $ 0.2431 $ \\
\hline
$SU(3)$ & Bonanno et al.~\cite{Bonanno:2020hht,Panagopoulos:2011rb,Bonati:2015sqt} & $ 0.0289(13) $ & $ 0.2222 $ \\
$SU(4)$ & Bonanno et al.~\cite{Bonanno:2020hht} & $ 0.02499(54) $ & $ 0.2344 $ \\
$SU(6)$ & Bonanno et al.~\cite{Bonanno:2020hht} & $ 0.02214(69) $ & $ 0.2431 $ \\
\hline
$SU(2)$ & Athenodorou et al.~\cite{Athenodorou:2021qvs} & $ 0.05565(64) $ & $ 0.1875 $ \\
$SU(3)$ & Athenodorou et al.~\cite{Athenodorou:2021qvs} & $ 0.0325(11) $ & $ 0.2222 $ \\
$SU(4)$ & Athenodorou et al.~\cite{Athenodorou:2021qvs} & $ 0.02469(67) $ & $ 0.2344 $ \\
$SU(5)$ & Athenodorou et al.~\cite{Athenodorou:2021qvs} & $ 0.0213(13) $ & $ 0.2400 $ \\
\hline
\end{tabular}
\end{table}

\begin{figure}[t]
\centering
\begin{picture}(300,230)
\put(0,0){\includegraphics[width=.45\textwidth]
{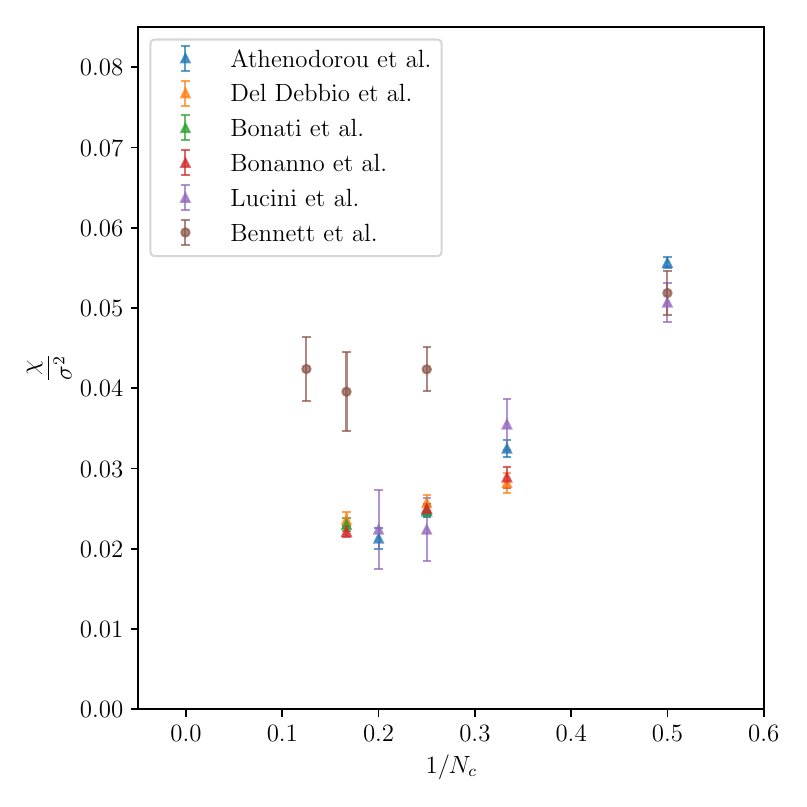}}
 \end{picture}
\caption{Topological susceptibility $\chi$,
in units of the string tension $\sigma$, in the continuum limit, for various groups $SU(N_c)$ and $Sp(N_c)$,
 and as a function of the parameter $1/N_c$. The measurements reported here are labelled by the collaboration 
 that published them, and  are also 
 summarised in Table~\ref{Fig:table}.\label{Fig:top}}
\end{figure}

\begin{figure}[t]
\centering
\begin{picture}(300,230)
\put(0,0){\includegraphics[width=.45\textwidth]
{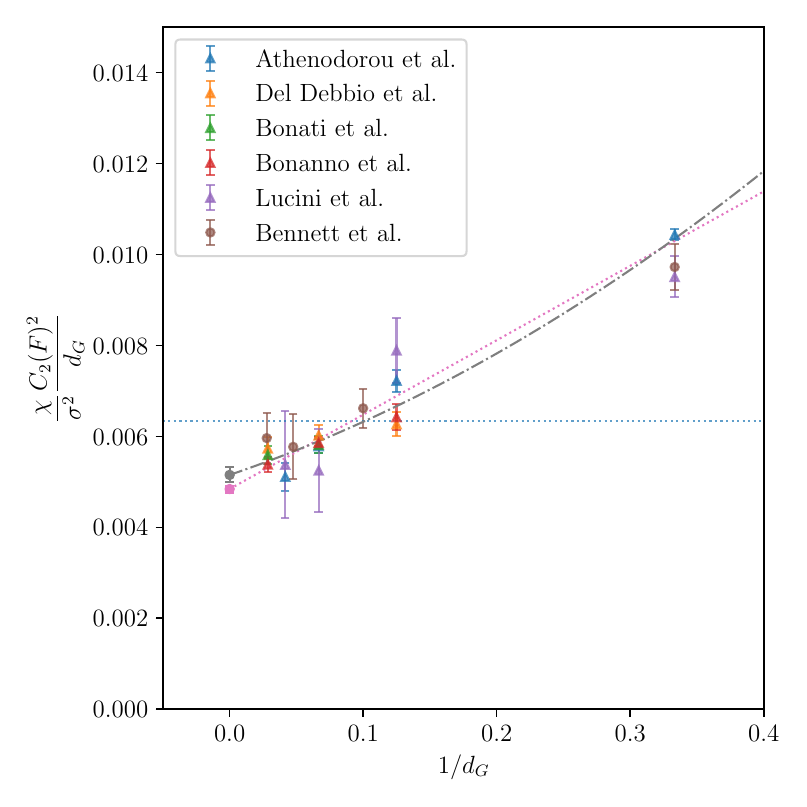}}
\end{picture}
\caption{Ratio
of topological susceptibility and string tension squared, rescaled  
by the group factor $C_2(F)^2/d_G$, as a function of
  $1/d_G$.
We also show the best-fit results of a 2-parameter fit (dotted line) and of a 3-parameter fit including $O(1/d_G^2)$
corrections (dashed line), as explained in the main text.
The horizontal dashed line is the NDA estimate $1/(4\pi)^2$.
\label{Fig:scaled}}
\end{figure}

We summarise in Table~\ref{Fig:table} lattice measurements
for the quantity $\chi/\sigma^2$ taken from 
Refs.~\cite{Lucini:2001ej,DelDebbio:2002xa,Bonati:2016tvi,Bonanno:2020hht,Athenodorou:2021qvs,Bennett:2022ftz,
Panagopoulos:2011rb, Bonati:2015sqt},
extrapolated to the continuum limit.
The same results are graphically displayed in Fig.~\ref{Fig:top}, where we organise
the measurements in terms of 
(the inverse of)
the number of colors $N_c$ in the gauge groups $SU(N_c)$ and $Sp(N_c)$, respectively.
In the table, we show also the group factor $C_F^2/d_G$,
which we use in Fig.~\ref{Fig:scaled} to rescale the measurements of $\chi/\sigma^2$, 
as described in Section~\ref{Sec:Nc}.
In this second plot we  also change  the abscissa to display $1/d_G$; 
for large $N_c$, $d_G\propto N_c^2$, 
and this more physical choice removes conventional ambiguities in comparing 
across different sequences of groups within Cartan's classification.
The data of Tab.~\ref{Fig:table} and the analysis code used to prepare
Figs.~\ref{Fig:top} and~\ref{Fig:scaled}, as well as the numbers quoted later in this Section,
are available at Ref.~\cite{datapackage, codepackage}.

Before proceeding, we comment on some subtleties about the
numerical results we quote, which have been obtained 
with heterogeneous treatments of systematic effects.
 The topological charge in pure gauge theories can be computed 
in different ways~\cite{Alexandrou:2017hqw},
 from ensembles of gauge configurations
generated with Monte Carlo algorithms,
all converging towards the same continuum limit.
Two  technical aspects deserve special attention.
Firstly, the continuum $\chi$ is related to the lattice $\chi_L$ by both 
additive and multiplicative
renormalisation. 
Second, the lattice discretisation renders the lattice topological charge, $Q_L$, non-integer.

All quoted calculations of $\chi$ make use of the 
 definition of $Q_L$ that employs the clover-leaf plaquette~\cite{Sheikholeslami:1985ij,Hasenbusch:2002ai}
on ensembles of configurations generated 
with the Cabibbo-Marinari implementation of the heat bath algorithm~\cite{Cabibbo:1982zn}.
In order to circumvent the noisy signal resulting from ultraviolet fluctuations of $Q_L$, 
one exploits the stability of the topological charge under smooth deformations of the fields,
 and computes it after a smoothing process such as cooling or Wilson flow. 
An integer value of $Q_L$ on the lattice can then be assigned either by 
 \emph{small-instanton-correction}~\cite{Lucini:2001ej}, or by
 \emph{correction-and-rounding}~\cite{DelDebbio:2002xa}. The former consists of
rounding the lattice topological charge to one of its neighbouring integer values, 
chosen with the
sign of the net contribution of small instantons. The latter comprises  rescaling $Q_L$
by minimising the average deviation of the lattice topological charge from integer multiples.

For $SU(N_c)$ gauge theories, Ref.~\cite{DelDebbio:2002xa} assigns
integer values to $Q_L$ by correction-and-rounding on cooled configurations
and computes the continuum limit of $\chi$ for $N_c=3,\,4,\,6$.
The same strategy  is used in Ref.~\cite{Bonati:2016tvi}, 
which reports the continuum limits for $N_c=4,\,6$.
With respect to these two
works, Ref.~\cite{Bonanno:2020hht} differs because the configurations
 are obtained by
an algorithm that considers a larger ensemble of systems with boundary conditions interpolating
 from periodic to open to soften the effects of topological freezing (see the quoted work for details);
the continuum limits are then obtained for $N_c=3,\,4,\,6$ although for $SU(3)$
the numerical results are taken from Refs.~\cite{Panagopoulos:2011rb, Bonati:2015sqt}.
By contrast, in Refs.~\cite{Lucini:2001ej,Athenodorou:2021qvs}  small-instanton-correction 
is applied to $Q_L$, obtained from cooled configurations, 
and the continuum $\chi$ is then extrapolated for $N_c=2,\,3,\,4,\,5$.

 In the case of $Sp(N_c)$ gauge theories, we borrow the
 results from a companion publication, Ref.~\cite{Bennett:2022ftz}, 
 which is part of the ongoing programme of study of $Sp(N_c)$
 lattice gauge theories~\cite{ 
 Bennett:2017kga,
 Bennett:2020qtj,
 Bennett:2019jzz,
 Bennett:2019cxd,
 Bennett:2022yfa}, 
 and uses the HiRep code~\cite{DelDebbio:2008zf},
  adapted to $Sp(N_c)$ groups~\cite{Bennett:2017kga}.
 The lattice topological charge is obtained from Wilson-flowed
configurations~\cite{Luscher:2010iy,Luscher:2013vga}, 
and correction-and-rounding is used to assign integer topological 
charge. The topological susceptibility $\chi$ is obtained in the continuum
 limit for $N_c=2,\,4,\,6,\,8$.

By comparing Figs.~\ref{Fig:top} and~\ref{Fig:scaled}, we observe two 
interesting facts. Firstly, the two sequences of measurements of $\chi/\sigma^2$
are clearly dissimilar, yet they share interesting properties at the extrema: measurements by different
collaborations for  $Sp(2)\sim SU(2)$ are in broad agreement, and going to large $N_c$ the 
two sequences show a tendency to converge towards two different constants for $N_c\gsim 4$.
Second, once we apply the rescaling by the group factor, $C_F^2/d_G$,
the two sequences can no longer be distinguished, the measurements 
for $Sp(N_c)$ and $SU(N_c)$ theories agreeing with one another, given current
uncertainties.
A rough estimate, based upon naive dimensional analysis (NDA)~\cite{Georgi:1992dw}, yields:
\beqs
\eta_{\chi}\,=\,\frac{\chi C_2(F)^2}{\sigma^2 d_G} &=&{\cal O}\left(\frac{1}{(4\pi)^2}\right).
\eeqs
This estimate falls straight  in the middle of the range of measurements,
possibly by  mere  numerical coincidence.
Yet, it is  remarkable that no more than a factor of $2$ separates 
existing measurements, for all groups $G$, and that this estimate yields the correct order of magnitude.

The scaling procedure allows us to perform a simple
global  fit of the whole set of measurement, in the form
\beqs
\eta_{\chi}\,=\,\frac{\chi C_2(F)^2}{\sigma^2 d_G}&=&a+\frac{c}{d_G}\,.
\label{Eq:fit}
\eeqs
The result of the fit, which has reduced 
 $\tilde{\mathcal{X}}^2\equiv{\mathcal{X}}^2/N_{\rm d.o.f.} =1.58$,
 is $a=0.004842(77)$ and $c=0.01635(46)$.
 Visual inspection of Fig.~\ref{Fig:scaled} and Table~\ref{Fig:table}
highlights  some modest tension between 
measurements performed by different collaborations for $SU(2)$, 
as well as for $SU(3)$, suggesting that for these two groups the systematic uncertainty 
is not negligible, compared to the statistical uncertainty. 
To  quantify this effect, we repeat the same 
fitting procedure, but by omitting the $Sp(N_c)$ measurements,
and obtain as a result that $\tilde{\mathcal{X}}^2 = 1.83$,
hence demonstrating that the combination of measurements taken in 
theories with the two families of groups does not
affect the goodness of the fit.

We also performed alternative fits, by including corrections  ${\cal O}(1/\sqrt{d_G})$ 
or ${\cal O}(1/d_G^2)$, to test the scaling hypothesis we made; these additional terms do not
change appreciably the results of the maximum likelihood analysis. Our final  result is
\beqs
\lim_{N_c\rightarrow \infty} 
\eta_{\chi}
= \left(48.42 \pm 0.77 \pm 3.31\right)\times 10^{-4}\,,
\label{Eq:fitres}
\eeqs
where the first error is the statistical one from the 2-parameter fit in the form Eq.~(\ref{Eq:fit}),
while the second  is the systematic error of the fitting procedure. The latter is
 conservatively estimated as 
the difference between using in the extrapolation either
 the 2-parameter fit or a 3-parameter fit including  
an additional term proportional to $1/d_G^2$---we
show the result of both fits in Fig.~\ref{Fig:scaled}.

 For $SU(N_c)$,  $C_2(F)^2\rightarrow d_G/4$ in the large-$N_c$ limit, hence our combined  result
in Eq.~(\ref{Eq:fitres}) can be recast as
$\chi/\sigma^2\rightarrow 0.01937\pm 0.00136$.
This is 
$\simeq 1.4$
standard deviations lower than the result 
$\chi/\sigma^2\rightarrow 0.0221(14)$ 
from Ref.~\cite{DelDebbio:2002xa},  but in excellent
agreement with Ref.~\cite{Bonanno:2020hht}, which quotes
$\chi/\sigma^2\rightarrow 0.0199(10)$,
and with Ref.~\cite{Athenodorou:2021qvs},
from which one deduces that   
$\chi/\sigma^2\rightarrow 0.01836(56)$.
%%%%%%%%%%%%%%%%%%%%%
%%%%%%%%%%%%%%%%%%%%%
\section{Outlook}
\label{Sec:outlook}

We proposed a rescaling  by group-theoretical factors
of the dimensionless quantity $\chi/\sigma^2$,
the ratio of topological susceptibility and square of the string tension,
to yield $\eta_{\chi}$, a quantity that can be
meaningfully  compared across different  (four-dimensional) Yang-Mills theories.
We collected from the literature the results of the continuum limit extrapolation
of several independent lattice measurements of $\eta_{\chi}$ in theories with  groups $SU(N_c)$ and $Sp(N_c)$.
All measurements of $\eta_{\chi}$ are of the order of magnitude indicated by 
a rough NDA estimate.
The  two sequences of groups display the same
functional dependence of $\eta_{\chi}$ on the dimension $d_G$ of the group,
in support of the proposed rescaling.
We assessed this statement by performing a combined fit of all the measurements,
and by extrapolating towards the large-$N_c$ limit.

We conclude by highlighting  a number of open questions, deserving of further future investigation.
The numerical evidence  we collected suggests that the group-theoretical scaling we proposed
allows  to combine measurements of $\chi$ within the sequences of
$SU(N_c)$ and $Sp(N_c)$ Yang-Mills theories. It would be fascinating to
extend this analysis to other choices of gauge group.
After rescaling, there remains clearly visible a non-trivial (though mild) dependence on the group dimension;
the precise functional form of the quantity $\chi C_2(F)^2/\sigma^2 d_G$ remains a subject for 
non-perturbative studies.
It would be interesting to
reassess these statements with future higher precision measurements.

%%%%%%%%%%%
%%%%%%%%%%%
%%%%%%%%%%%

%%%%%%%%%%%%%%%%%%%%%%%%%%%%%%%%%%%%%%%%
\vspace{1.0cm}
\begin{acknowledgments}

The work of EB has been funded in part by the Supercomputing Wales project, 
which is part-funded by the European Regional Development Fund (ERDF) via Welsh Government,
and by the UKRI Science and Technology Facilities Council (STFC)
 Research Software Engineering Fellowship EP/V052489/1

The work of DKH was supported by the National Research Foundation of Korea (NRF) grant funded by the Korea government (MSIT) (2021R1A4A5031460) and also by Basic Science Research Program through the National Research Foundation of Korea (NRF) funded by the Ministry of Education (NRF-2017R1D1A1B06033701).

The work of JWL is supported by the National Research Foundation of Korea (NRF) grant funded 
by the Korea government(MSIT) (NRF-2018R1C1B3001379). 

The work of CJDL is supported by the Taiwanese MoST grant 109-2112-M-009-006-MY3. 

The work of BL and MP is supported in part by the STFC 
Consolidated Grants No. ST/P00055X/1 and No. ST/T000813/1.
 BL and MP received funding from
the European Research Council (ERC) under the European
Union’s Horizon 2020 research and innovation program
under Grant Agreement No.~813942. 
The work of BL is further supported in part 
by the Royal Society Wolfson Research Merit Award 
WM170010 and by the Leverhulme Trust Research Fellowship No. RF-2020-4619.

The work of DV is supported in part by the INFN HPC-HTC project 
and in part by the Simons Foundation under the program “Targeted 
Grants to Institutes” awarded to the Hamilton Mathematics Institute.

Numerical simulations have been performed on the 
Swansea University SUNBIRD cluster (part of the Supercomputing Wales project) and AccelerateAI A100 GPU system,
on the local HPC
clusters in Pusan National University (PNU) and in National Yang Ming Chiao Tung University (NYCU),
and the DiRAC Data Intensive service at Leicester.
The Swansea University SUNBIRD system and AccelerateAI are part funded by the European Regional Development Fund (ERDF) via Welsh Government.
The DiRAC Data Intensive service at Leicester is operated by 
the University of Leicester IT Services, which forms part of 
the STFC DiRAC HPC Facility (www.dirac.ac.uk). The DiRAC 
Data Intensive service equipment at Leicester was funded 
by BEIS capital funding via STFC capital grants ST/K000373/1 
and ST/R002363/1 and STFC DiRAC Operations grant ST/R001014/1. 
DiRAC is part of the National e-Infrastructure.

\vspace{1.0cm}

{\bf Open Access Statement - } For the purpose of open access, the authors have applied a Creative Commons 
Attribution (CC BY) licence  to any Author Accepted Manuscript version arising.

\end{acknowledgments}
%%%%%%%%%%%%%%%
%%%%%%%%%%%%%%% Appendix
%%%%%%%%%%%%%%%
\FloatBarrier
\appendix
%%%%%%%%%%%%
%%%%%%%%%%%% Bibliography
%%%%%%%%%%%%

\end{document}